\documentclass[twocolumn,aps,prc,superscriptaddress,preprintnumbers,floatfix,nofootinbib]{revtex4}
\usepackage{epsfig,graphics}
\usepackage{graphicx}
\usepackage{dcolumn,multirow}
\usepackage{bm}
\usepackage{amsmath}
\usepackage[usenames]{color}
\usepackage{ulem} 
\newcommand{\er}{$\pm$}

\begin{document}

\title{\boldmath Is there a narrow $N(1685)$\,?}

\newcommand*{\HISKP}{Helmholtz--Institut f\"ur Strahlen--
                     und Kernphysik, Universit\"at Bonn, 53115 Bonn, Germany}
\newcommand*{\GATCHINA}{National Research Centre ``Kurchatov Institute'',
                     Petersburg Nuclear Physics Institute, Gatchina, 188300 Russia}
\newcommand*{\JLAB}{Thomas Jefferson National Accelerator Facility, Newport News, Virginia 23606, USA}

\affiliation{\HISKP}
\affiliation{\GATCHINA}
\affiliation{\JLAB}

\author{A.V. Anisovich} \affiliation{\HISKP} \affiliation{\GATCHINA}
\author{V.~Burkert} \affiliation{\JLAB}
\author{E.~Klempt}  \affiliation{\HISKP} \affiliation{\JLAB}
\author{V.A.~Nikonov} \affiliation{\HISKP} \affiliation{\GATCHINA}
\author{A.V.~Sarantsev} \affiliation{\HISKP} \affiliation{\GATCHINA}
\author{U.~Thoma} \affiliation{\HISKP}

\date{\today}

\begin{abstract}
The helicity-dependent observable $E$ for the reaction $\gamma d\to \eta n (p)$ with a spectator
proton was recently measured by the A2 Collaboration at MAMI in Mainz. The data were interpreted
as further evidence for a narrow resonance with spin and parity $J^P=1/2^+$ ($P_{11}$ wave). However,
a full partial wave analysis without any narrow resonance
leads to an excellent description of the data; imposing a narrow resonance with the
properties suggested by the A2 Collaboration leads to a significant deterioration of the fit
quality: there is no need for a narrow resonance.

\end{abstract}

\pacs{25.75.-q}

\maketitle
A narrow structure was observed at a mass of about 1685\,MeV in the $\gamma d\to \eta n (p)$ excitation function
\cite{Kuznetsov:2006kt,Jaegle:2008ux,Kuznetsov:2010as,Jaegle:2011sw,Werthmuller:2013rba,%
Werthmuller:2014thb,Kuznetsov:2015nla}.
The structure was interpreted \cite{Polyakov:2003dx,N_1680} as the nonstrange member of the
antidecuplet of pentaquarks with spin-parity
$J^P=1/2^+$ predicted by Diakonov, Petrov, and Polyakov \cite{Diakonov:1997mm}. In 2012, the observations reported
in~\cite{Kuznetsov:2006kt,Kuznetsov:2010as,Jaegle:2011sw} were introduced into the Review of Particle Properties (RPP)
under the heading of a new one-star nucleon resonance $N(1685)$~\cite{Beringer:1900zz} but
was removed from the listings in the most recent issue of RPP \cite{Olive:2016xmw}.
The interpretation of the structure as narrow resonance was supported by further studies
\cite{Arndt:2003ga,Choi:2006,Fix:2007,Shrestha:2012}, the results reported 
in~\cite{Anisovich:2008wd} were ambiguous.

However, also coupled-channel and interference effects of known nucleon resonances
have been discussed in the literature to explain the narrow structure. The Gie\ss en
group interpreted the narrow dip in the $\gamma d\to \eta n (p)$ excitation function as
$N(1650)1/2^-$ and $N(1710)1/2^+$ coupled-channel effect~\cite{Shklyar:2006xw}, Shyam and 
Scholten assign the
dip to interference effects between the $N(1650)1/2^-$, $N(1710)1/2^+$, and $N(1720)3/2^+$
resonances~\cite{Shyam:2008fr}; alternatively, the dip could be produced to effects from
strangeness threshold openings \cite{Doering:2009qr}.

The narrow dip can, however, also be explained naturally by interference effects in the 
$J^P=1/2^-$ wave \cite{Anisovich:2008wd,Zhong:2011ti,Anisovich:2013jya,Anisovich:2015tla}. 
In \cite{Anisovich:2015tla}, the precise data
reported by the A2 Collaboration at MAMI~\cite{Werthmuller:2013rba,Werthmuller:2014thb} were
used to study the structure. It was found that it can be explained quantitatively by interference
of the two nucleon resonances $N(1535)1/2^-$ and $N(1650)1/2^-$ within the $J^P=1/2^-$ partial wave.
Fits which included a narrow $J^P=1/2^+$ resonance returned a zero production strength. If the properties
of the narrow $J^P=1/2^+$ resonance as reported in~\cite{Werthmuller:2013rba,Werthmuller:2014thb}
were imposed, the fit deteriorated significantly.

\begin{figure}
\begin{center}
\begin{tabular}{c}
\hspace{-1mm}\includegraphics[width=0.48\textwidth]{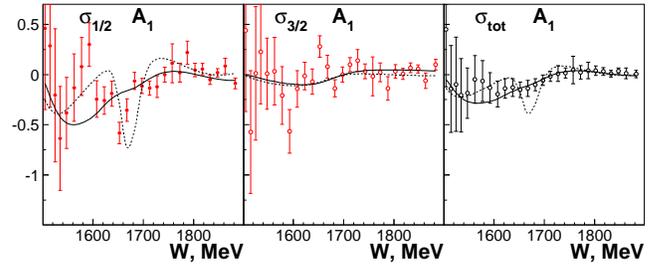}
\end{tabular}\vspace{-3mm}
\end{center}
\caption{Legendre coefficients of the angular distributions of $\sigma_{1/2}$, $\sigma_{3/2}$
\cite{Witthauer:2016mpi}, and  $\sigma_{\rm tot}$ for the reaction $\gamma d\to \eta n (p)$ 
where $\sigma_{\rm tot}$ is calculated as $(\sigma_{1/2}+\sigma_{3/2})/2$. The experimental results
(red circles) are compared to a BnGa fit without a narrow resonance (solid curve) or a fit 
imposing a narrow resonance (dotted curve). }
\label{L1}\vspace{-1mm}
\end{figure}

Recently, the A2 Collaboration at MAMI reported a measurement of the 
helicity-dependent double polarization variable $E$ of the
$\gamma d\to \eta n (p)$ reaction \cite{Witthauer:2016mpi} where
$E= (\sigma_{1/2}-\sigma_{3/2})/(\sigma_{1/2}+\sigma_{3/2})$, with
$\sigma_{h}$ being the cross section for $\gamma d\to \eta n (p)$ with neutron and
photon spin aligned (helicity $h=3/2$) and or opposite ($h=1/2$).  The data show clearly that the structure
originates from the $h=1/2$ contribution. The authors fitted the angular distributions
(five data points per energy interval) with third-order Legendre polynomial functions and found
a narrow dip at 1650\,MeV in the first order Legendre coefficient. They concluded:
{\it The extracted Legendre coefficients of the angular distributions for $\sigma_{1/2}$ are in
good agreement with recent reaction model predictions assuming a narrow resonance
in the $P_{11}$ wave as the origin of this structure.} In this paper we will show that 
their conclusions are incompatible with the data.

As a first step, we repeated the fit with Legendre polynomials. Figure~\ref{L1} shows the first-order
Legendre coefficients A$_1^{\sigma_{1/2}}$, A$_1^{\sigma_{3/2}}$, and A$_1^{\sigma_{\rm tot}}$
as functions of the $n\eta$ invariant mass for fits to the angular distributions of $\sigma_{1/2}$,
$\sigma_{3/2}$, and  $\sigma_{\rm tot}=(\sigma_{1/2}+ \sigma_{3/2})/2$. The coefficients 
A$_0^{\sigma_{1/2}}$, A$_0^{\sigma_{3/2}}$, and A$_0^{\sigma_{\rm tot}}$ are similar to 
the corresponding total cross sections, the coefficients A$_2$ and A$_3$ for the cross sections
$\sigma_{1/2}$ and $\sigma_{3/2}$ are shown in \cite{Witthauer:2016mpi}. In the coefficient 
A$_1^{\sigma_{1/2}}$ there is indeed a narrow dip at about 1650\,MeV. 
Since the $J^P=1/2^-$ partial wave
dominates the reaction, significant contributions to A$_1^{\sigma_{1/2}}$ have to come from the
interference between the $J^P=1/2^-$ partial wave and $P$-wave contributions. Indeed,
a comparison of  A$_1^{\sigma_{1/2}}$ with fit results shows that a model assuming no 
$N(1685)$ (Fig.~\ref{L1}, solid curve) does not reproduce the narrow dip while a model
which includes a narrow $N(1685)$ (Fig.~\ref{L1}, dotted curve) gives qualitative 
agreement between data and prediction. These observations are the basis for the 
conjecture in \cite{Witthauer:2016mpi} that a narrow resonance has been observed.
\begin{figure}[pt]
\begin{center}
\includegraphics[width=0.48\textwidth,height=0.46\textwidth]{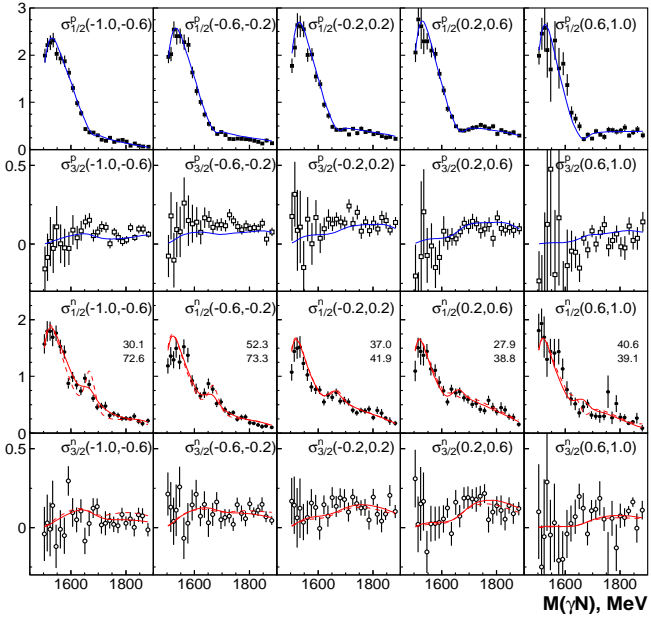}\vspace{-4mm}
\end{center}
\caption{Excitation functions $\sigma_{1/2}$ and $\sigma_{3/2}$ for 5 bins
in $\cos\theta^*_\eta$ for the reaction $\gamma d\to\eta p(n)$ (top 2 rows)
and $\gamma d\to\eta n(p)$ (bottom 2 rows) and new BnGa fits. The data are 
from \cite{Witthauer:2016mpi}, statistical and systematic errors are added quadratically.
The solid lines represent the BnGa fit without an additional narrow resonance, the 
dashed lines a fit in which a narrow resonance is imposed with the
properties given in \cite{Werthmuller:2013rba}. The two numbers give the $\chi^2$
contribution of the bin, the upper number without, the lower number including
the narrow resonance.}
 \label{fig:diffp}
\end{figure}
There are, however, a few arguments which disagree with this conjecture.

\begin{figure}[pt]
\begin{center}
\begin{tabular}{cc}
             \includegraphics[width=0.24\textwidth,height=0.22\textwidth]{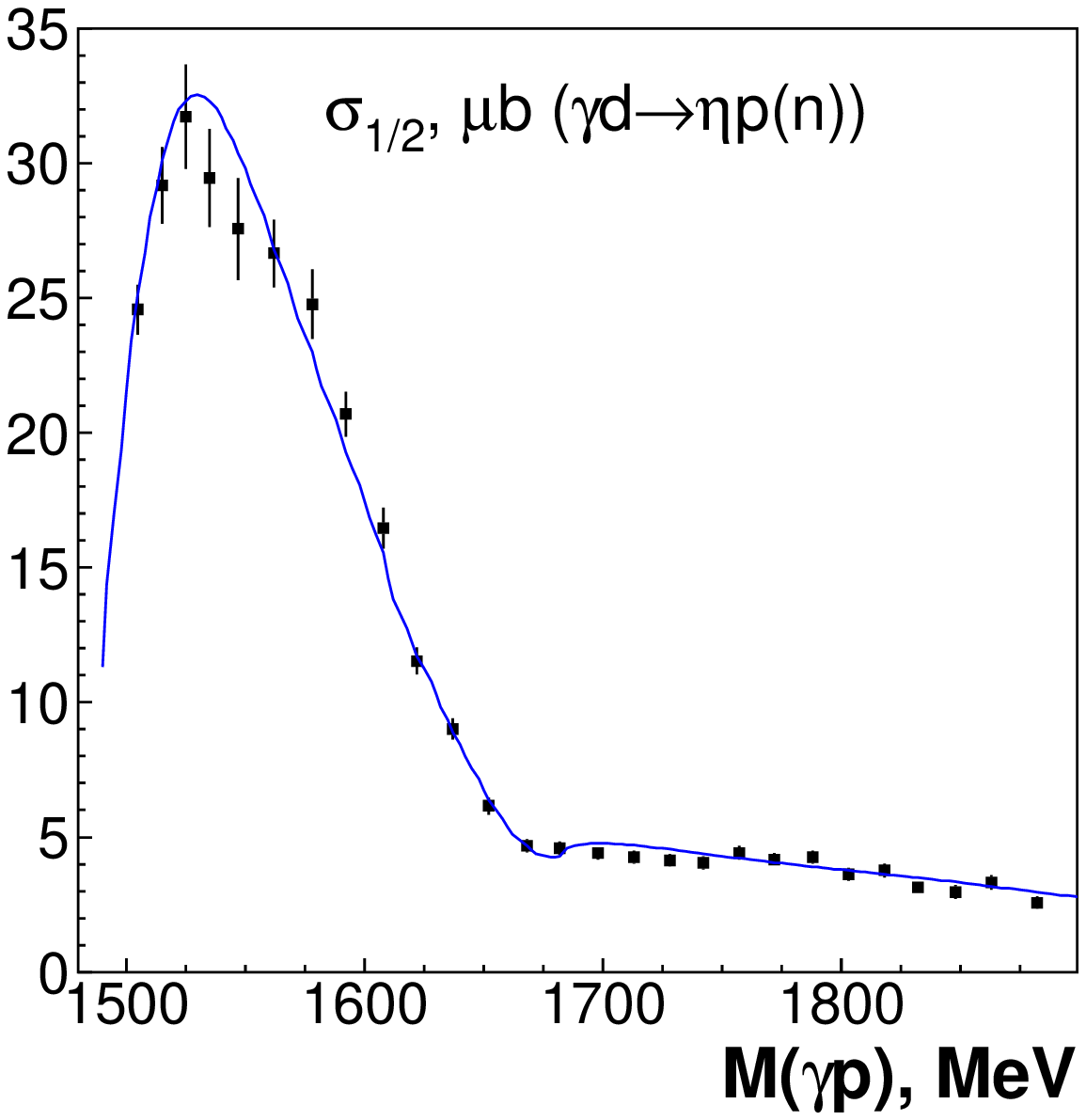}&
\hspace{-1mm}\includegraphics[width=0.24\textwidth,height=0.22\textwidth]{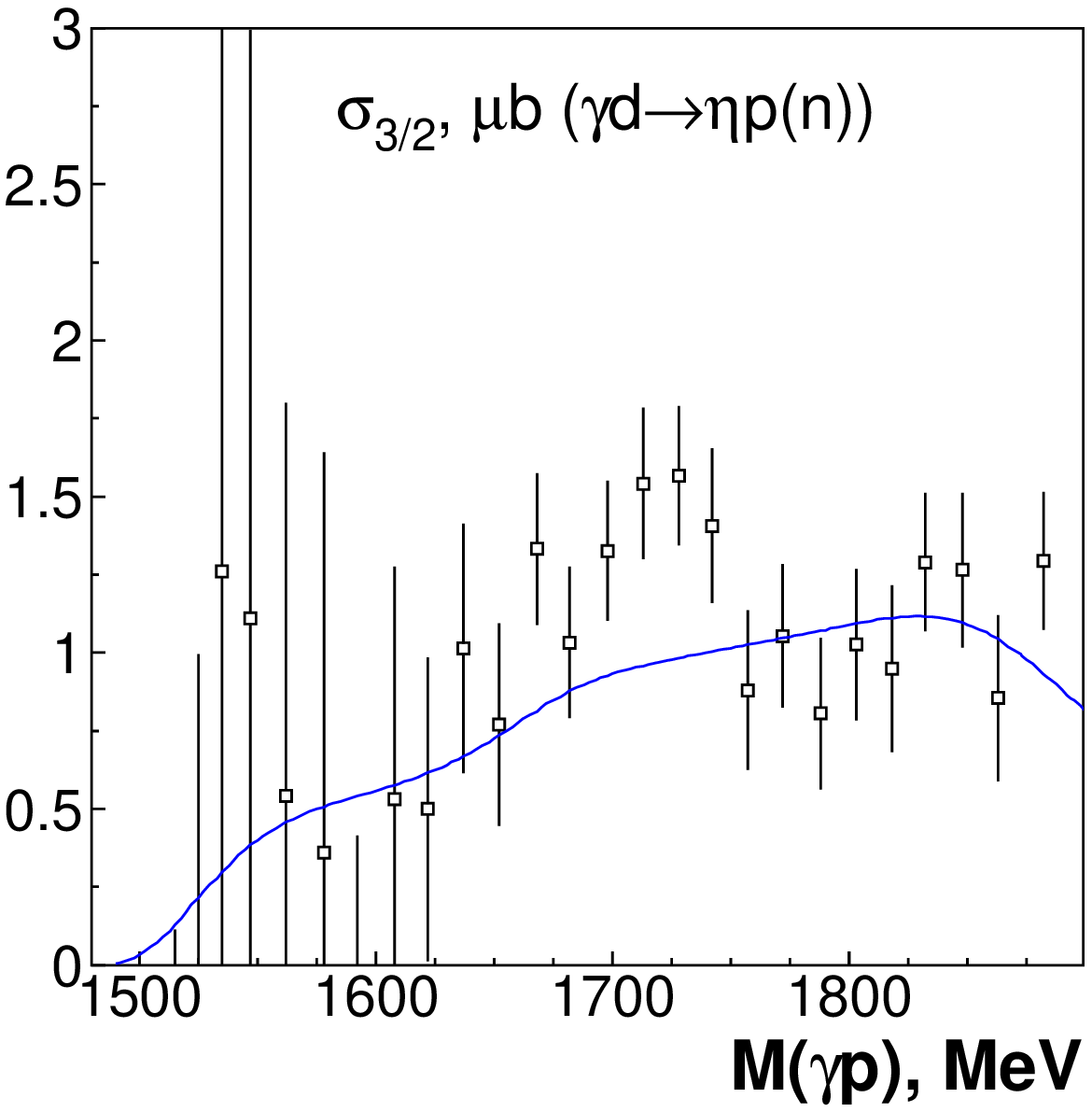}\\[1ex]
             \includegraphics[width=0.24\textwidth,height=0.22\textwidth]{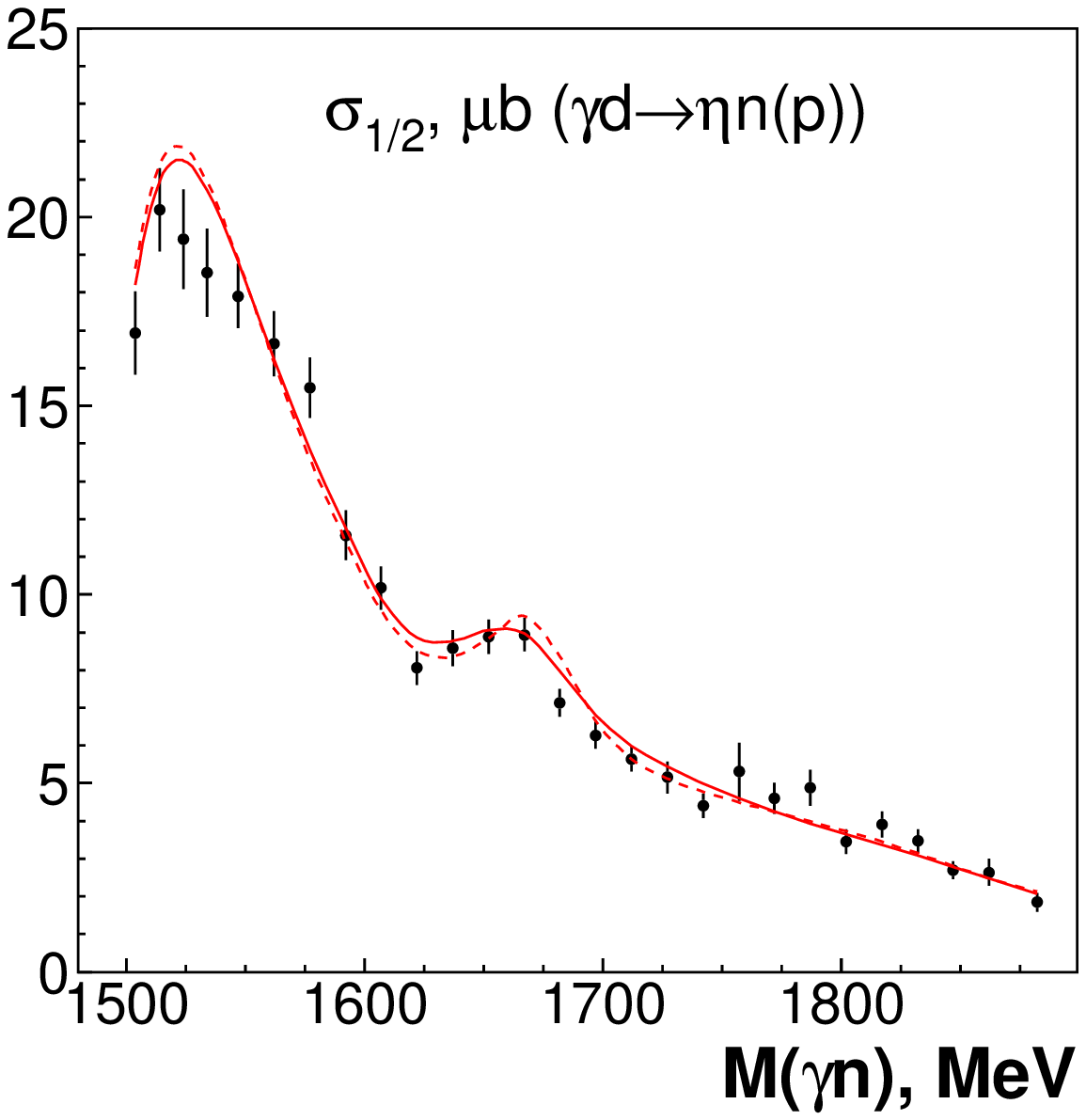}&
\hspace{-1mm}\includegraphics[width=0.24\textwidth,height=0.22\textwidth]{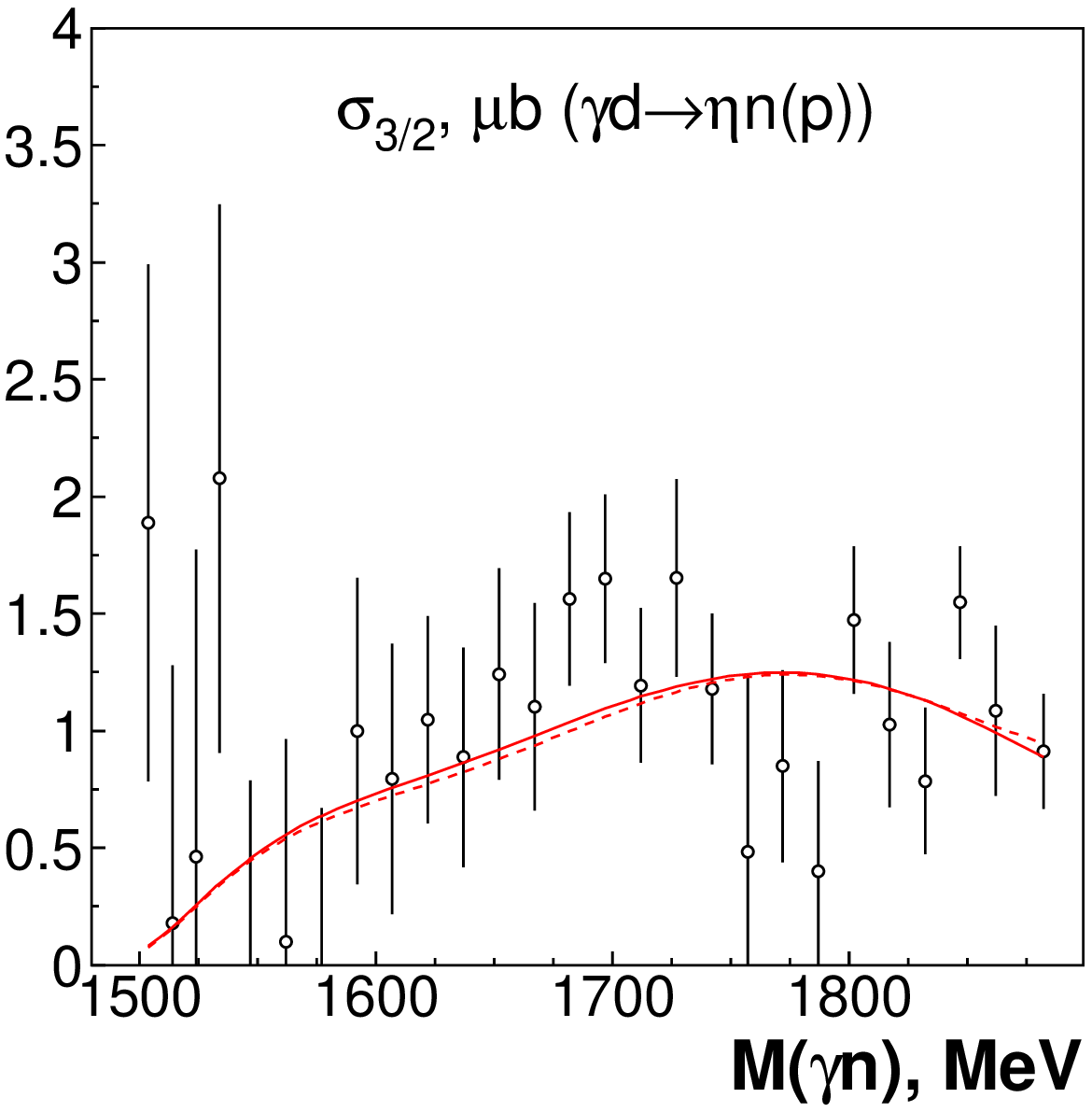}\vspace{-2mm}
\end{tabular}
\end{center}
\caption{The total cross sections $\sigma_{1/2}$ , $\sigma_{3/2}$ for the reactions
$\gamma d\to\eta p(n)$ (top) and $\gamma d\to\eta n(p)$ (bottom) and new BnGa fits. 
The solid lines represent a fit without a narrow resonance, the dashed lines in the 
bottom figure represent a fit in which a narrow resonance is imposed with the properties 
given in~\cite{Werthmuller:2013rba}. \vspace{-2mm}}
 \label{fig:totp}
\end{figure}
\noindent

The dip in A$_1^{\sigma_{1/2}}$ is statistically significant. Relative to the solid line (representing the fit
with no $N(1685)$), the dip in A$_1^{\sigma_{1/2}}$ has a mean deviation $-0.24\pm 0.04$ and
contributes $\chi^2=15.9$ for two data
points.
\begin{table*}[!]
\caption{\label{helamp}Helicity amplitudes determined from a fit without a narrow $N(1685)$ resonance.
The T-matrix couplings are the quantities which are listed in the RPP; K-matrix couplings are given in addition. The new results are compared to those obtained in \cite{Anisovich:2015tla} which
are listed in small numbers. The comparison shows the impact of the new data from \cite{Witthauer:2016mpi}
and \cite{Mueller:2017spl}.
}
\begin{center}
\renewcommand{\arraystretch}{1.5}
\begin{tabular}{|l|cccl|l|cccl|}
\hline\hline
&& $N(1535)1/2^-$ & $N(1650)1/2^-$ && && $N(1535)1/2^-$ & $N(1650)1/2^-$ &\\\hline
 \multirow{5}{*}{\boldmath$p$}&\underline{T-matrix}&   $0.093\pm 0.009$ &   $0.032\pm 0.006$ &GeV$^{-1/2}$ &
 \multirow{5}{*}{\boldmath$n$}&\underline{T-matrix}&  -$0.088\pm 0.004$ &   $0.016\pm 0.004$ &GeV$^{-1/2}$\\[-1ex]
 &\scriptsize \cite{Anisovich:2015tla}&\scriptsize   $0.114\pm 0.008$ & \scriptsize   $0.032\pm 0.007$ &\scriptsize  GeV$^{-1/2}$ &
 &\scriptsize \cite{Anisovich:2015tla}&\scriptsize   -$0.095\pm 0.006$& \scriptsize   $0.019\pm 0.006$ &\scriptsize  GeV$^{-1/2}$\\
 &\underline{Phase}&    8\er4$^\circ$ &  7\er7$^\circ$  & &  &\underline{Phase}&   5\er4$^\circ$ &  -28\er10$^\circ$  &\\[-1ex]
 &\scriptsize \cite{Anisovich:2015tla}& \scriptsize   10\er5$^\circ$ &\scriptsize  -2\er11$^\circ$  & &
&\scriptsize \cite{Anisovich:2015tla}&\scriptsize    8\er5$^\circ$ &\scriptsize   0\er15$^\circ$  &\\[-1ex]
&\underline{K-matrix} & $0.112\pm 0.008$ & $0.075\pm 0.006$  & & &\underline{K-matrix}&  -$0.160\pm 0.030$       & -$0.052\pm 0.005$   &\\[-1ex]
&\scriptsize \cite{Anisovich:2015tla}& \scriptsize  $0.096\pm 0.007$ &\scriptsize  $0.075\pm 0.007$  & &
&\scriptsize \cite{Anisovich:2015tla}
& \scriptsize  -$0.120\pm 0.006$       &\scriptsize  -$0.052\pm 0.006$   &\\
\hline\hline
\end{tabular}\vspace{-3mm}
\end{center}
\end{table*}
However, there is a peak in A$_1^{\sigma_{3/2}}$ as well, at the same mass and of similar size and shape
as the dip in A$_1^{\sigma_{1/2}}$. The peak deviates from the solid line by 
$+0.25\pm 0.04$, contributes $\chi^2=12.7$, and is thus of similar importance as the dip. The
coefficient A$_1^{\sigma_{\rm tot}}$ follows precisely the fit with no
$N(1685)$, the data are compatible with the fit, with $\chi^2=2.1$ for the two data points.
If the dip in A$_1^{\sigma_{1/2}}$ had a physical significance, it should be seen in
A$_1^{\sigma_{\rm tot}}$ with a strength as given by the dotted line. But it is not.
There is hence the suspicion that the dip might be a statistical fluctuation: a small change in
the observable $E$ may lead to a disappearance of the dip and the peak.

To test this hypothesis, we performed overall fits. In these fits most particle properties 
are frozen to the values derived from fits to pion and photo-induced reactions
off protons. For $\gamma n$ reactions we use the data listed in \cite{Anisovich:2015tla} and,
in addition, the new MAMI data~\cite{Witthauer:2016mpi}. The latter data are shown in 
Fig.~\ref{fig:diffp} and Fig.~\ref{fig:totp}, the solid line is our fit without introduction of
a narrow resonance.  
For the differential cross sections from \cite{Werthmuller:2013rba,Werthmuller:2014thb} and
\cite{Witthauer:2016mpi}, the fit returns a $\chi^2_{\rm MAMI}= 1205$ for 1150
data points. Obviously, there is no need to introduce $N(1685)$. When $N(1685)$ was
enforced in the fit with properties as given in \cite{Werthmuller:2013rba}, i.e. with $M=1670$ MeV,
width $\Gamma= 30$ MeV,
and $\sqrt{Br(\eta n)}A_n^{1/2}=\tilde a$ [GeV$^{-\frac12}$\,10$^{-3}$] = $12.3$
[GeV$^{-\frac12}$\,10$^{-3}$], the fit returned $\chi^2_{\rm MAMI}=1834$ for the
1150 data points from \cite{Werthmuller:2013rba,Werthmuller:2014thb,Witthauer:2016mpi}.
The $\chi^2$'s for the new data from \cite{Witthauer:2016mpi} are shown in Fig.~\ref{fig:diffp} 
for each angular bin of $\sigma_{1/2}$ for $\gamma d\to \eta n (p)$, the sum is
$\chi^2=187.9$ for the fit without narrow resonance and 265.8 when it is imposed.

\begin{figure}[pt]
\begin{center}
\includegraphics[width=0.25\textwidth,height=0.22\textwidth]{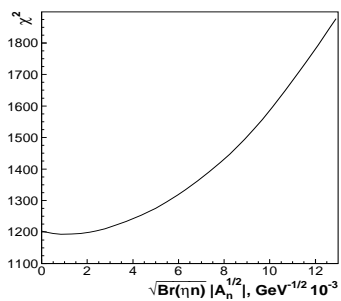}
\end{center}
\caption{\label{chi2}Increase of $\chi^2$ when a narrow resonance with mass $M=1670$ MeV,
width $\Gamma= 30$ MeV, and $J^P=1/2^+$ is imposed
as a function of the signal strength $\tilde a = \sqrt{Br(\eta n)}A_n^{1/2}$. The number
of data points is 1150.}
\end{figure}

If the production strength is fitted freely, it reduced to $1.2$\,[GeV$^{-\frac12}$\,10$^{-3}$] and
the total $\chi^2$ improved by 12 units to 1193.
This production strength corresponds to a contribution which is
about 100 times smaller than the contribution claimed in \cite{Werthmuller:2013rba,Werthmuller:2014thb}.
Fig.~\ref{chi2} shows how the $\chi^2$ increases with the strength of an imposed narrow
$N(1670)$.

The new data on $E$ for the reaction $\gamma d\to  \eta p (n)$ -- with a spectator
neutron -- in \cite{Witthauer:2016mpi}
differed significantly from first BnGa fits which were performed before the data on
double-polarization observables on $\gamma p\to \eta p$ on protons became 
available~\cite{Mueller:2017spl}. 
To explore this discrepancy, we included the new MAMI data for
$\eta$ production off protons (bound in deuterons) \cite{Witthauer:2016mpi} in the fits. 
Figures \ref{fig:diffp}
and \ref{fig:totp} show that the new data can be included in the fit without any problems,
after a slight tuning of the parameters. In Table~\ref{helamp}, we show the helicity amplitudes
obtained in the new fit in comparison to the fit presented in \cite{Anisovich:2015tla}. The changes
in the photocouplings of $N(1535)1/2^-$ and $N(1650)1/2^-$ for protons are likely due to the inclusion
of the new data on $\gamma p\to \eta p$ \cite{Mueller:2017spl}.

Summarizing, we have studied the new data on the helicity dependence of the reaction
$\gamma d\to \eta n (p)$ with a spectator proton measured by the A2 Collaboration at
MAMI in Mainz \cite{Witthauer:2016mpi}. We cannot confirm the conclusions of the authors
that the dip in the first-order Legendre coefficient in an expansion of the angular
distributions of $\sigma_{1/2}$ is due to a narrow $J^P=1/2^+$ resonance. First, the
dip is accompanied by a peak in the first-order Legendre coefficient of $\sigma_{3/2}$
of the same shape suggesting that the dip is due to a statistical fluctuation in the
measurement of $E$. Second, a partial wave analysis
without a narrow $J^P=1/2^+$ resonance is excellent, the inclusion of it with the
reported properties leads to a significantly worse description of the data.

{\small Comments of Bernd Krusche and Volker Metag to an early version of this
comment are kindly recognized.
The work was supported by the \textit{Deutsche Forschungsgemeinschaft} (SFB/TR110),
the \textit{U.S. Department of Energy} (DE-AC05-06OR23177),
and the \textit{Russian Science Foundation} (RSF 16-12-10267).}

\bibliographystyle{apsrev}

\end{document}